Research Article

The Missing Link of Jewish European Ancestry: Contrasting the Rhineland and the Khazarian Hypotheses

Eran Israeli-Elhaik[1,2]

[1] Department of Mental Health, Johns Hopkins University Bloomberg School of Public Health, Baltimore, MD, USA, 21208.

[2] McKusick-Nathans Institute of Genetic Medicine, Johns Hopkins University School of Medicine, Baltimore, MD, USA, 21208.

Please address all correspondence to Eran Elhaik at eelhaik@jhsph.edu

Phone: 410-502-5740. Fax: 410-502-7544.

**Abstract**

The question of Jewish ancestry has been the subject of controversy for over two centuries and has yet to be resolved. The "Rhineland Hypothesis" proposes that Eastern European Jews emerged from a small group of German Jews who migrated eastward and expanded rapidly. Alternatively, the "Khazarian Hypothesis" suggests that Eastern European descended from Judean tribes who joined the Khazars, an amalgam of Turkic clans that settled the Caucasus in the early centuries CE and converted to Judaism in the 8$^{th}$ century. The Judaized Empire was continuously reinforced with Mesopotamian and Greco-Roman Jews until the 13$^{th}$ century. Following the collapse of their empire, the Judeo-Khazars fled to Eastern Europe. The rise of European Jewry is therefore explained by the contribution of the Judeo-Khazars. Thus far, however, their contribution has been estimated only empirically; the absence of genome-wide data from Caucasus populations precluded testing the Khazarian Hypothesis. Recent sequencing of modern Caucasus populations prompted us to revisit the Khazarian Hypothesis and compare it with the Rhineland Hypothesis. We applied a wide range of population genetic analyses — including principal component, biogeographical origin, admixture, identity by descent, allele sharing distance, and uniparental analyses — to compare these two hypotheses. Our findings support the Khazarian Hypothesis and portray the European Jewish genome as a mosaic of Caucasus, European, and Semitic ancestries, thereby consolidating previous contradictory reports of Jewish ancestry.

**Introduction**

Contemporary Eastern European Jews comprise the largest ethno-religious aggregate of modern Jewish communities, accounting for nearly 90% of over 13 million Jews worldwide (United Jewish Communities 2003). Speculated to have emerged from a small Central European founder group and maintained high endogamy, Eastern European Jews are considered invaluable subjects in disease studies (Carmeli 2004), although their ancestry remains debatable among geneticists, historians, and linguists (Wexler 1993; Brook 2006). Because correcting for population structure and using suitable controls are critical in medical studies, it is vital to test the different hypotheses pertaining to explain the ancestry of Eastern European Jews. One of the major challenges for any hypothesis is to explain the massive presence of Jews in Eastern Europe, estimated at eight million people at the beginning of the 20$^{th}$ century. The two dominant hypotheses depict either a sole Middle Eastern ancestry or a mixed Middle Eastern-Caucasus-European ancestry.

The "Rhineland Hypothesis" envisions modern European Jews to be the descendents of the Judeans – an assortment of Israelite-Canaanite tribes of Semitic origin (Figures 1,2) (Supplementary Note). It proposes two mass migratory waves: the first occurred over the next two hundred years after the Muslim conquest Palestine (638 CE) and consisted of devoted Judeans who left Muslim Palestine for Europe (Dinur 1961; Sand 2009). It is unclear whether these migrants joined the existing Judaized Greco-Roman communities and the extent of their contribution to the Southern European gene pool. The second wave occurred at the beginning of the 15$^{th}$ century by a group of 50,000 German Jews who migrated eastward and ushered an apparent hyper-baby-boom era for half a millennia affecting only Eastern Europe Jews (Atzmon et al. 2010). The annual growth rate that accounted for the populations' rapid expansion from this small group was estimated at 1.7-2% (Straten 2007), twice the rate of any documented baby-boom period and lasting 20 times longer. This growth rate is also one order of magnitude larger than that of Eastern European non-Jews in the 15$^{th}$-17$^{th}$ centuries. The Rhineland Hypothesis predicts a Middle Easter ancestry to European Jews and high genetic similarity among European Jews (Ostrer 2001; Atzmon et al. 2010; Behar et al. 2010).

The competing "Khazarian Hypothesis" considers Eastern European Jews the descendants of ancient and late Judeans who joined the Khazars, a confederation of Slavic, Scythian, Sabirs, Finno-Ugrian, Alan, Avars, Iranian, and Turkish tribes who formed in the northern Caucasus one of most powerful and pluralistic empires during the late Iron Age and converted to Judaism in the 8$^{th}$ century CE (Figures 1-2) (Polak 1951; Brook 2006; Sand 2009). The Khazarian, Armenian, and Georgian populations forged from this amalgamation of tribes (Polak 1951), followed by high levels of isolation, differentiation, and genetic drift *in situ* (Balanovsky et al. 2011). The population structure of the Judeo-Khazars was further reshaped by multiple migrations of Jews from the Byzantine Empire and Caliphate to the Khazarian Empire (Figure 1). The collapse of the Khazar Empire followed by the Black Death (1347-1348) accelerated the progressive depopulation of Khazaria (Baron 1993) in favor of the rising Polish Kingdom and Hungary (Polak 1951). The newcomers mixed with the existing Jewish communities established during the uprise of Khazaria and spread to Central and Western Europe. The Khazarian Hypothesis predicts that European Jews comprise of Caucasus, European, and Middle Eastern ancestries and is distinct from the Rhineland Hypothesis in the existence of a large genetic signature of Caucasus populations. Because some Eastern European Jews migrated west and admixed with the neighboring Jewish and non-Jewish populations they became distinct from the remaining Eastern European Jews. Therefore, different European Jewish communities are expected to be heterogeneous. Alternative hypotheses, such as the "Greco-Roman Hypothesis" (Zoossmann-Diskin 2010), were also proposed to explain the origins of European Jews; however, they do not explain the massive presence of Eastern Europeans Jews in the 20$^{th}$ century and therefore were not tested here.

Many genetic studies attempting to settle these competing hypotheses yielded inconsistent results. Some studies pointed at the genetic similarity between European Jews and Middle

Eastern populations such as Palestinians (Hammer et al. 2000; Nebel et al. 2000; Atzmon et al. 2010), while few pointed at the similarity to Caucasus populations like Adygei (Behar et al. 2003; Levy-Coffman 2005; Kopelman et al. 2009), and others pointed at the similarity to Southern European populations like Italians (Atzmon et al. 2010; Zoossmann-Diskin 2010). Most of these studies were done in the pre-genomewide era, using uniparental markers and included different reference populations, which makes it difficult to compare their results. More recent studies employing whole genome data reported high genetic similarity of European Jews to Druze, Italian, and Middle Eastern populations (Atzmon et al. 2010; Behar et al. 2010).

Motivated by the recent availability of genome-wide data for key populations, the current study aims to uncover the ancestry of Eastern and Central European Jews by contrasting the Rhineland and Khazarian Hypotheses. These hypotheses were tested here by comparing the biogeographical genetic profile of European Jews with indigenous Middle Eastern and Caucasus populations using a wide set of population genetic tools including principal component analyses (PCAs), identification of the biogeographical origins of European Jews, admixture, identity by descent (IBD), allele sharing distance (ASD), and finally by comparing haplogroup frequencies of Y and mtDNA.

As the Judeans and Khazars have been vanquished and their remains have yet to be sequenced, in accordance with previous studies (Levy-Coffman 2005; Kopelman et al. 2009; Atzmon et al. 2010; Behar et al. 2010), contemporary Middle Eastern and Caucasus populations were used as surrogates. Palestinians were considered Proto-Judeans because they shared a similar linguistic, ethnic, and geographic background with the Judeans and were shown to share common ancestry with European Jews (Bonné-Tamir and Adam 1992; Nebel et al. 2000; Atzmon et al. 2010; Behar et al. 2010). Similarly, Caucasus Armenian and Georgians were considered Proto-Khazars because they emerged from the same cohort as the Khazars (Polak 1951; Dvornik 1962; Brook 2006). Mitochondrial DNA (mtDNA) analyses corroborated the genetic similarity between Armenians and Georgians (Behar et al. 2010) and estimated their approximate divergence time from the Turks and Iranians 600 and 360 generations ago, respectively (Schonberg et al. 2011).

Although both the Rhineland and Khazarian Hypotheses depict a Judean ancestry and are not mutually exclusive, they are well distinguished, as Caucasus and Semitic populations are considered ethnically and linguistically distinct (Patai and Patai 1975; Wexler 1993; Balanovsky et al. 2011). Jews, according to either hypothesis, are an assortment of tribes who accepted Judaism and maintained it up to this date and are, therefore, expected to exhibit certain heterogeneity with their neighboring populations. Because, according to both hypotheses, Eastern European Jews arrived in Eastern Europe roughly at the same time (13$^{th}$ and 15$^{th}$ centuries), we assumed that they experienced similar low and fixed admixture rates with the neighboring populations, estimated at 0.5% per generation over the past 50 generations (Ostrer 2001). These relatively recent admixtures have likely reshaped the population structure of all European Jews and increased the genetic distances from the Caucasus or Middle Eastern populations. Therefore, we do not expect to achieve perfect matching with the surrogate populations but rather to estimate their relatedness.

**Materials and Methods**

**Data analysis**. The complete data set contained 1,287 unrelated individuals of 8 Jewish and 74 non-Jewish populations genotyped over 531,315 autosomal single nucleotide polymorphisms (SNPs). An LD-pruned data set was created by removing one member of any pair of SNPs in strong LD ($r^2$>0.4) in windows of 200 SNPs (sliding the window by 25 SNPs at a time) using indep-pairwise in PLINK (Purcell et al. 2007). This yielded a total of 221,558 autosomal SNPs that were chosen for all autosomal analyses except the IBD analysis that utilized the complete data set. Both data sets were obtained from http://www.evolutsioon.ut.ee/MAIT/jew_data/ (Behar et al. 2010). MtDNA and Y chromosomal data were obtained from previously published data sets as appear in Behar et al. (2010). These markers were chosen to match the phylogenetic level of resolution achieved in previously reported data sets and represent a diversified set of markers. A total of 6,089 and 5,303 samples were assembled for mtDNA and Y chromosomal analyses, respectively, from 27 populations (Tables S1, S2).

In common parlance, Eastern and Central European Jews are practically synonymous with Ashkenazi Jews and are considered a single entity (Tian et al. 2008; Atzmon et al. 2010; Behar et al. 2010). However, the term is misleading, for the Hebrew word "Ashkenaz" was applied to Germany in mediaeval rabbinical literature - thus contributing to the narrative that modern Eastern European Jewry originated on the Rhine. We thus refrained from using the term "Ashkenazi Jews." Jews were roughly subdivided into Eastern (Belorussia, Latvia, Poland, and Romania) and Central (Germany, Netherland, and Austria) European Jews. In congruence with the literature that considers "Ashkenazi Jews" distinct from "Sephardic Jews," we excluded the later. Small populations (<7 samples) were excluded from the PC and IBD analyses. Complete population notation is described in Table S3.

**Principal component analysis (PCA)**. PCA calculations were carried out using smartpca of the EIGENSOFT package (Patterson, Price, and Reich 2006). To circumvent biases caused by unequal number of populations (Price et al. 2006; McVean 2009), we developed a simple dual-PCA framework consisting of three "outgroup" populations that are available in large sample sizes and are the least admixed: Mbuti and Biaka Pygmies (Africa), French Basques (Europe), and Han Chinese (Asia) and two populations of interest — all of equal sample sizes. The cornerstone of this framework is that it minimizes the number of significant PCs to four or fewer (Tracy-Widom test, $p<0.01$) and maximizes the portion of explained variance to over 20% for the first two PCs. Convex hulls were calculated using Matlab "convhull" function and plotted around the cluster centroids. Relatedness between two populations of interest was estimated by the commensurate overlap of their clusters.

**Estimating the biogeographical origins of population**. To decrease the bias caused by multiple populations of uneven size (Patterson, Price, and Reich 2006; McVean 2009), we used the dual-PCA framework of three outgroup populations and two populations of interest: a population of known geographical origin during the relevant time period shown to cluster with the population of interest and the population in question. The first four populations were used as a training set for the population in question. PCA calculations were carried out as described above. The rotation angle of PC1-PC2 coordinates was calculated as described by Novembre et al. (2008). Briefly, in each figure the PC axes were rotated to find the angle that maximizes the summed correlation of the median PC1 and PC2 values of the training populations with the latitude and longitude of their countries. Latitudinal and longitudinal data were obtained from the literature or by the country's center. Geodesic distances were calculated in kilometers using the Matlab function "distance."

**Admixture analysis**. A structure-like approach was applied in a *supervised* learning mode as implemented in ADMIXTURE (Alexander, Novembre, and Lange 2009). ADMIXTURE provides an estimation of the individual's ancestries from the allele frequencies of the designated ancestral populations. ADMIXTURE's bootstrapping procedure with default parameters was used to calculate the standard errors. Populations were sorted by their mean African and Asian ancestries.

**Identity by descent (IBD) analysis.** To detect IBD segments we ran fastIBD ten times using different random seeds and combined the results as described by Browning and Browning (2011). Segments were considered to be IBD only if the fastIBD score of the combined analysis was less than $e^{-10}$. This low threshold corresponds to long shared haplotypes (≥1 cM) that are likely to be IBD. Short Gaps (<50 indexes) separating long domains were assumed to be false-negative and concatenated. Pairwise-IBD segments between European Jews and different populations were obtained by finding the maximum total IBD sharing between each European Jew and all other individuals of a particular population.

**Allele sharing distances (ASD)**. ASD was used for measuring genetic distances between populations as it is less sensitive to small sample sizes than other methods. Pairwise ASD was calculated using PLINK (Purcell et al. 2007), and the average ASD between populations *I* and *J*, was computed as:

$$W^{IJ} = (\sum_{i \in I} \sum_{i \in J} W^{ij})/nm,$$

where $W^{ij}$ is the distance between individuals *i* and *j* from populations *I* and *J* of sizes *n* and *m*, respectively. To verify that these ASD differences are significant, a bootstrap approach was used with the null hypothesis: $H_0$: ASD ($p_1$, $p_2$) = ASD ($p_1$, $p_3$), where the ASD between populations $p_1$ and $p_2$ is compared to the ASD between populations $p_2$ and $p_3$. To compare continental Jewish communities, individuals were grouped by their continent and the comparison was carried as described. Similarly to Behar et al. (2010), we adopted a bootstrap procedure of 1000 repeats accounting for variance resulting from both sample and site selection to determine the significance of differences in ASD in each row in Table 1 (calculated separately for Jewish and non-Jewish communities) (Tables S4-S6). The same approach was also used to calculate the standard errors in ASD estimates (Table S7).

**Uniparental analysis.** To infer the migration patterns of European Jews, we integrated haplogroup data from over 11,300 uniparental chromosomes with geographical data. The haplogroup frequencies were then compared between populations to obtain a measure of distance between populations. Pairwise genetic distances between population haplogroups (Tables S1-S2) were estimated by applying the Kronecker function as implemented in Arlequin version 3.1 (Excoffier, Laval, and Schneider 2005). In brief, similarity between populations was defined as the fraction of haplogroups that the two populations share as measured by the Kronecker function $\delta_{xy}(i)$:

$$d_{xy} = \sum_{i=1}^{L} \delta_{xy}(i), \tag{1}$$

which equals 1 if the haplogroup frequency of the *i*-th haplogroup is non-zero for both populations and equals 0 otherwise. In other words, populations sharing the same exact haplogroups or their mutual absence are considered more genetically similar than populations with different haplogroups. For brevity, we considered only haplogroups with frequencies higher than 0.5%. This measure has several desirable properties that make it an excellent measure for estimating genetic distance between populations, such as a simple interpretation in terms of homogeneity and applicability to both mtDNA and Y chromosomal data.

**Results**

PCA was used to identify independent dimensions that capture most of the information in the data. Although PCA has many attractive properties, it should be practiced with caution to circumvent biases due to the choice of populations and varying sample sizes (Price et al. 2006; McVean 2009). We applied PCA using two frameworks: the "multi-population" carried for all populations (Figure 3) and separately for Eurasian populations along with Pygmies and Han Chinese (Figure S2) and our novel "dual-population" consisting of three "outgroup" populations and two populations of interest (Figure S3). In all analyses, the studied samples aligned along the two well-established geographic axes of global genetic variation: PC1 (sub-Saharan Africa versus the rest of the Old World) and PC2 (east versus west Eurasia) (Li et al. 2008). Our results reveal geographically refined groupings, such as the nearly symmetrical continuous European rim extending from West to Eastern Europeans, the parallel Caucasus rim, and the Near Eastern populations (Figure S1) organized in Turk-Iranian and Druze clusters (Figure 3). Middle Eastern populations form a gradient along the diagonal line between Bedouins and Near Eastern populations that resembles their geographical distribution. The remaining Egyptians and the bulk of Saudis distribute separately from Middle Eastern populations.

European Jews are expected to cluster with native Middle Eastern or Caucasus populations according to the Rhineland or Khazarian Hypotheses, respectively. The results of all PC analyses (Figures 3, S2-3) show that over 70% of European Jews and almost all Eastern European Jews cluster with Armenian, Georgian, and Azerbaijani Jews within the Caucasus rim (Figures 3, S3). Nearly 15% of Central European Jews cluster with Druze and the rest cluster with Cypriots. All European Jews cluster distinctly from the Middle Eastern cluster. A strong evidence for the Khazarian Hypothesis is the clustering of European Jews with the populations that reside on opposite ends of ancient Khazaria: Armenians, Georgians, and Azerbaijani Jews (Figure 1). Because Caucasus populations remained isolated in the Caucasus region and because there are no records of Caucasus populations mass-migrating to Eastern and Central Europe prior to the fall of Khazaria (Balanovsky et al. 2011), these findings imply a shared origin for European Jews and Caucasus populations.

To locate the biogeographical origin of a population, Novembre et al. (2008) proposed a PCA-based approach, accurate to a few hundred kilometers. We implemented this approach using a dual-PCA framework to reduce sample biases. To demonstrate the accuracy of this approach, we first sought to identify the biogeographical origin of Druze (Hitti 1928). We traced Druze biogeographical origin to the geographical coordinates: 38.6±3.45° N, 36.25±1.41° E (Figure S4) in the Near East (Figure S1). Half of the Druze clustered tightly in Southeast Turkey, and the remaining was scattered along northern Syria and Iraq. These results are in agreement with Shlush et al.'s (2008) findings using mtDNA anaysis. The inferred geographical positions were used in the subsequent analyses.

European Jews should be positioned within the Middle East according to the Rhineland Hypothesis or in the region between the Near East and the Caucasus according to the Khazarian Hypotheses. Though the geographical origins of European Jews varied for different reference populations (Figure 4, S5), the results converged to Southern Khazaria along modern Turkey, Armenia, Georgia, and Azerbaijan (Figure 1). The smallest deviations in the geographical coordinates were obtained with Armenians for both Eastern (38±2.7° N, 39.9±0.4° E) and Central (35±5° N, 39.7±1.1° E) European Jews (Figure 4). Remarkably, the mean coordinates of Eastern European Jews are 560 kilometers from Khazaria's southern border (42.77° N, 42.56° E) near Samandar – the capital city of Khazaria from 720 to 750 CE (Polak 1951). Eastern European Jews clustered tightly compared to Central European Jews in all analyses.

The duration, direction, and rate of gene flow between populations determine the proportion of admixture and the total length of chromosomal segments that are identical by descent (IBD). Admixture calculations were carried out using a *supervised* learning approach in a structure-like analysis. This approach has many advantages over the *unsupervised* approach that not only traces ancestry to *K* abstract unmixed populations under the assumption that they evolved independently (Chakravarti 2009; Weiss and Long 2009) but also problematic when applied to study Jewish ancestry, which can be dated as far back as 3,000 years (Figure 2). Admixture was calculated with a reference set of seven populations representing genetically distinct regions: Pygmies (Africa), French Basque (West Europe), Chuvash (East Europe), Han Chinese (Asia), Palestinians (Middle East), Turk-Iranians (Near East), and Armenians (Caucasus) (Figure 5). The Khazarian Hypothesis predicts that European Jews share similar ancestry with Caucasus, European, and Middle Eastern populations, whereas the Rhineland Hypothesis predicts their sharing similar ancestry with native Middle Eastern populations. The ancestral components grouped all populations by their geographical regions with European Jews cluster with Caucasus populations. As expected, Eastern and Western European ancestries exhibit opposite gradients among European populations. The Near Eastern-Caucasus ancestries are dominant among Central (38%) and Eastern (32%) European Jews followed by Western European ancestry (30%). Among non-Caucasus populations the Armenian ancestry is the largest among European Jews (26%) and Cypriots (31%). These populations also exhibit the largest faction of Palestinian ancestry among non-Middle Eastern populations. As both Armenian and Palestinian are absence in Eastern European populations, our findings suggest that Eastern European Jews acquired these ancestries prior to their arrival to Eastern Europe. Although the Rhineland Hypothesis explains the Palestinian ancestry by stating that Jews migrated from Palestine to Europe in the $7^{th}$ century, it fails to explain the large Armenian ancestry which is nearly endemic to Caucasus populations.

Although they clustered with Caucasus populations (Figure 5), Eastern and Central European Jews share a large fraction of Western European and Middle Eastern ancestries, both absent in Caucasus populations, excepting Armenians who share 15±2% Middle Eastern ancestry (Figure S6). According to the Khazarian Hypothesis, the Western European ancestry was imported to Khazaria by Greco-Romans Jews and, whereas the Middle Eastern ancestry point to the contribution of both Judeans from Mesopotamian Jews (Polak 1951; Koestler 1976; Sand 2009) (Figure 1). Central and Eastern European Jews differ mostly in their Middle Eastern (30% and 25%, respectively) and Eastern European ancestries (3% and 12%, respectively), probably due to late admixture.

Druze exhibit a large Turkic ancestry (83%) in accordance with their Near Eastern origin (Figure S4). Druze and Cypriot appear similar to European Jews in their Middle Eastern and Western European ancestries though they differ largely in the proportion of Caucasus ancestry. These results can explain the genetic similarity between European Jews, Southern Europeans, and Druze reported in studies that excluded Caucasus populations (Price et al. 2008; Atzmon et al. 2010; Zoossmann-Diskin 2010). Overall, our results portray the European Jewish genome as a mosaic of Caucasus, Western European, Middle Eastern, and Eastern European ancestries in decreasing proportions.

To glean further details of the genomic regions contributing to the genetic similarity between European Jews and the perspective populations, we compared their total genomic regions shared by IBD. If European Jews emerged from Caucasus populations the two would share longer IBD regions than with Middle Eastern populations. The IBD analysis exhibits a skewed bimodal distribution embodying a major Caucasus ancestry with a minor Middle Eastern ancestry (Figure 6), consistent with the admixture results (Figure 5). The total IBD regions shared between European Jews and Caucasus populations (9.5 cM on average) are significantly larger than regions shared with Palestinians (5.5 cM) (Kolmogorov-Smirnov goodness-of-fit test, $p<0.001$). To the best of our knowledge, these are the largest IBD regions ever reported between European

Jews and non-Jewish populations. The decrease in total IBD between European Jews and other populations combined with the increase in distance from the Caucasus support the Khazarian Hypothesis.

Eastern European Jews are one of the most studied ethnicities, particularly in disease studies, due to the presumption of their high endogamy across all their communities. Consequently, under the Rhineland Hypothesis, both Eastern and Central European Jews are genetically more distant from their neighboring populations as compared to Middle Eastern populations. By contrast, the Khazarian Hypothesis depicts different Jewish communities as highly heterogeneous created by a mixture of mostly Caucasus populations along with Central European and Middle Eastern populations followed by admixture with the neighboring populations. We tested the level of Jews' endogamy by estimating the allele sharing distance (ASD) between Eurasian Jewish communities. We also compared the ASD distances between Jews and their non-Jewish neighbors, Caucasus, and Middle Eastern populations.

Our results expand the previous report of high endogamy in Jewish populations (Behar et al. 2010) and narrow the endogamy to regional Jewish communities (Table 1, left panel). Jews are significantly more similar to members of their own community than to other Jewish populations ($P<0.01$, bootstrap $t$-test), with the conspicuous exception of Bulgarian, Turkish, and Georgian Jews. These results stress the high heterogeneity among Jewish communities across Eurasia and even within communities, as in the case of the Balkan and Caucasus Jews.

When compared to non-Jewish populations, all Jewish communities were significantly ($P<0.01$, bootstrap $t$-test) distant from Middle Eastern populations and, with the exception of Central European Jews, significantly closer to Caucasus populations (Table 1, right panel). Similar findings were reported by Behar et al. (2010), although they were considered a bias in the calculations. The close genetic distance between Central European Jews and Southern European populations can be attributed to a late admixture. The results are consistent with our previous findings (Figure 5) in support of Khazarian Hypothesis. As the only commonality among all Jewish communities is their dissimilarity from Middle Eastern populations (Table 1), grouping different Jewish communities without correcting for their country of origin, as is commonly done, would increase their genetic heterogeneity.

Finally, we carried uniparental analyses on mtDNA and Y-chromosome comparing the haplogroup frequencies between European Jews and other populations. The Rhineland Hypothesis depicts Middle Eastern origins for European Jews' both paternal and maternal ancestries, whereas the Khazarian Hypothesis depicts a Caucasus ancestry along with a Southern European and Near Eastern contributions of migrates from Byzantium and the Caliphate, respectively. Because Judaism was maternally inherited only since the $3^{rd}$ century CE (Patai and Patai 1975), the mtDNA is expected to show a stronger local female-biased founder effect compared to the Y-chromosome. Haplogroup similarities between European Jews and other populations were plotted as heat maps on the background of their geographical locations (Figure 7). The pairwise distances between all studied populations are shown in Figure S7.

Our results shed light on sex-specific processes that, although not evident from the autosomal data, are analogous to those obtained from the biparental analyses. Both mtDNA and Y-chromosomal analyses yield high similarities between European Jews and Caucasus populations rooted in the Caucasus (Figure 7), in support of the Khazarian Hypothesis. Interestingly, the maternal analysis depicts a specific Caucasus founding lineage with a weak Southern European ancestry (Figure 7a), whereas the paternal ancestry reveals a dual Caucasus-Southern European origin (Figure 7b). As expected, the maternal ancestry exhibits a higher relatedness scale with narrow dispersal compared with the paternal ancestry.

Dissecting uniparental haplogroups allow us to delve further into European Jews' migration routes. Because the data do not specify whether the Southern Europe-Caucasus migration was ancient or recent nor indicate the migration's direction i.e., from Southern Europe to the Caucasus or the opposite, there are four possible scenarios. Of these, the only historically supportable scenarios are ancient migrations from Southern Europe toward Khazaria ($6^{th}$-$13^{th}$ centuries) and more recent migrations from the Caucasus to Central and Southern Europe ($13^{th}$-$15^{th}$ centuries) (Polak 1951; Patai and Patai 1975; Straten 2003; Brook 2006; Sand 2009). A westward migration from the diminished Khazaria toward Central and Southern Europe would have exhibited a gradient from the Caucasus toward Europe for both matrilineal and patrilineal lines. Such gradients were not observed. By contrast, Judaized Greco-Roman male-driven migration directly to Khazaria is consistent with historical demographic migrations and could have created the observed pattern. Moreover, we found little genetic similarity between European Jews and populations eastward to the Caspian Sea and southward to the Black Sea, delineating the geographical boundaries of Khazaria (Table 1, Figure 1).

**Discussion**

Eastern and Central European Jews comprise the largest group of contemporary Jews, accounting for nearly 90% of over 13 million worldwide Jews (United Jewish Communities 2003). Eastern European Jews made over 90% of European Jews before World War II. Despite of their controversial ancestry, European Jews are an attractive group for genetic and disease studies due to their presumed genetic history (Ostrer 2001). Correcting for population structure and using suitable controls are critical in disease studies, thus it is vital to determine whether European Jews are of Semitic, Caucasus, or other ancestry.

Though Judaism was born encased in theological-historical myth, no Jewish historiography was produced from the time of Josephus Flavius ($1^{st}$ century CE) to the $19^{th}$ century (Sand 2009). Early German historians bridged the historical gap simply by linking modern Jews directly to the ancient Judeans (Figure 1); a paradigm that was quickly embedded in medical science and crystallized as a narrative. Many have challenged this narrative (Koestler 1976; Straten 2007), mainly by showing that a sole Judean ancestry cannot account for the vast population of Eastern European Jews in the beginning of the 20$^{th}$ century without the major contribution of Judaized Khazars and by demonstrating that it is in conflict with anthropological, historical, and genetic evidence (Dinur 1961; Patai and Patai 1975; Baron 1993).

With uniparental and whole genome analyses providing ambiguous answers (Levy-Coffman 2005; Atzmon et al. 2010; Behar et al. 2010), the question of European Jewish ancestry remained debated mainly between the supporters of the Rhineland and Khazarian Hypotheses. The recent availability of genomic data of Caucasus populations (Behar et al. 2010) allowed testing the Khazarian Hypothesis for the first time and prompted us to contrast the Rhineland and Khazarian Hypotheses. To evaluate the two hypotheses, we carried out a series of comparative analyses between European Jews and surrogate Khazarian and Judean populations posing the same question each time: are Eastern and Central European Jews genetically closer to Caucasus or Middle Eastern populations? We emphasize that these hypotheses are not exclusive and that some European Jews may have other ancestries.

Our PC, biogeographical estimation, admixture, IBD, ASD, and uniparental analyses were consistent in depicting a Caucasus ancestry for European Jews. Our first analyses revealed tight genetic relationship of European Jews and Caucasus populations and pinpointed the biogeographical origin of European Jews to the south of Khazaria (Figures 3,4). Our later analyses yielded a complex multi-ethnical ancestry with a slightly dominant Near Eastern-Caucasus ancestry, large Southern European and Middle Eastern ancestries, and a minor Eastern European contribution; the latter two differentiated Central and Eastern European Jews (Figures 4, 5 and Table 1). While the Middle Eastern ancestry faded in the ASD and uniparental analyses, the Southern European ancestry was upheld probably attesting to its later time period (Table 1 and Figure 7).

We show that the Khazarian Hypothesis offers a comprehensive explanation to the results, including the reported Southern European (Atzmon et al. 2010; Zoossmann-Diskin 2010) and Middle Eastern ancestries (Nebel et al. 2000; Behar et al. 2010). By contrast, the Rhineland

Hypothesis could not explain the large Caucasus component in European Jews, which is rare in Non-Caucasus populations (Figure 5) and the large IBD regions shared between European Jews and Caucasus populations attesting to their common origins. A major difficulty with the Rhineland Hypothesis, in addition to the lack of historical and anthropological evidence to the multi-migration waves from Palestine to Europe (Straten 2003; Sand 2009), is to explain the vast population expansion of Eastern European Jews from 50 thousand (15$^{th}$ century) to 8 million (20$^{th}$ century). This growth could not possibly be the product of natural population expansion (Koestler 1976; Straten 2007), particularly one subjected to severe economic restrictions, slavery, assimilation, the Black Death and other plagues, forced and voluntary conversions, persecutions, kidnappings, rapes, exiles, wars, massacres, and pogroms (Koestler 1976; Sand 2009). Such an unnatural growth rate (1.7-2% annually) over half a millennia, affecting only Jews residing in Eastern Europe is commonly explained by a miracle (Atzmon et al. 2010). Unfortunately, this divine intervention explanation poses a new kind of problem - it is not science. Our findings reject the Rhineland Hypothesis and uphold the thesis that Eastern European Jews are Judeo-Khazars in origin. Further studies are necessary to confirm the magnitude of the Khazars demographic contribution to the demographic presence of Jews in Europe (Polak 1951; Dinur 1961; Koestler 1976; Baron 1993; Brook 2006).

The most parsimonious explanation for these findings is that Eastern European Jews are of Judeo-Khazarian ancestry forged over many centuries in the Caucasus. Jewish presence in the Caucasus and later Khazaria was recorded as early as the late centuries BCE and reinforced due to the increase in trade along the Silk Road (Figure 1), the decline of Judah (1$^{st}$-7$^{th}$ centuries), and the uprise of Christianity and Islam (Polak 1951). Greco-Roman and Mesopotamian Jews gravitating toward Khazaria were also common in the early centuries and their migrations were intensified following the Khazars' conversion to Judaism (Polak 1951; Brook 2006; Sand 2009). The eastward male-driven migrations (Figure 7) from Europe to Khazaria solidified the exotic Southern European ancestry in the Khazarian gene pool, (Figure 5) and increased the genetic heterogeneity of the Judeo-Khazars. The religious conversion of the Khazars encompassed all the Empire's citizens and subordinate tribes and lasted for the next 400 years (Polak 1951; Baron 1993) until the invasion of the Mongols (Polak 1951; Dinur 1961; Brook 2006). At the final collapse of their empire (13th century), the Judeo- Khazars fled to Eastern Europe and later migrated to Central Europe and admixing with the neighboring populations.

Historical and archeological findings shed light on the demographic events followed the Khazars' conversion. During the half millennium (740–1250 CE) of their existence, the Judeo-Khazars sent offshoots into the Slavic lands, such as Romania and Hungary (Baron 1993), planting the seeds of a great Jewish community to later rise in the Khazarian diaspora. We hypothesize that the settlement of Judeo-Khazars in Eastern Europe was achieved by serial founding events, whereby populations expanded from the Caucasus into Eastern and Central Europe by successive splits, with daughter populations expanding to new territories following changes in socio-political conditions (Gilbert 1993). As a result, the Jewish communities along the Caucasus borders appear more heterogeneous than other Jewish communities (Table 1), assuming an even and low admixture rate.

After the decline of their Empire, the Judeo-Khazars refugees sought shelter in the emerging Polish Kingdom and other Eastern European communities, where their expertise in economics, finances, and politics were valued. Prior to their exodus, the Judeo-Khazar population was estimated to be half a million in size, the same as the number of Jews in the Polish-Lithuanian kingdom four centuries later (Polak 1951; Koestler 1976). Some Judeo-Khazars were left behind, mainly in the Crimea and the Caucasus, where they formed Jewish enclaves surviving into modern times. One of the dynasties of Jewish princes ruled in the 15th century under the tutelage of the Genovese Republic and later of the Crimean Tartars. Another vestige of the Khazar nation are the "Mountain Jews" in the North Eastern Caucasus (Koestler 1976). In the 16th century the total Jewish population of the world amounted to about one million, suggesting that during the Middle Ages the majority of Jews were Judeo-Khazars in origin (Polak 1951; Koestler 1976).

The remarkable close proximity of European Jews and populations residing on the opposite ends of ancient Khazaria, such as Armenians, Georgians, Azerbaijani Jews, and Druze (Figures 3, S2-3, 5), supports a common Near Eastern-Caucasus ancestry. These findings are not explained by the Rhineland Hypothesis and are staggering due to the uneven demographic processes these populations experienced in the past eight centuries. The high genetic similarity between European Jews and Armenian compared to Georgians (Figures 5 and 6, Table 1) is particularly bewildering because Armenians and Georgians are very similar populations that share a similar genetic background (Schonberg et al. 2011) and long history of cultural relations (Payaslian 2007). We identified a small Middle Eastern ancestry in Armenians that does not exist in Georgians and is likely responsible to the high genetic similarity between Armenians and European Jews (Figure S6). Because the Khazars blocked the Arab approach to the Caucasus, we suspect that this ancestry was introduced by the Judeans arriving at a very early date to Armenia and were absorbed into the populations, whereas Georgian Jewry remained distinct (Shapira 2007). Similarly, the relatedness between European Jews and Druze reported here and in the literature (Behar et al. 2010) is explained by Druze Turkish-Southern Caucasus origins. Druze migrated to Syria, Lebanon, and eventually to Palestine between the $11^{th}$ and $13^{th}$ centuries during the Crusades, a time when the Jewish population in Palestine was at minimum. The genetic similarity between European Jews and Druze therefore supports the Khazarian Hypothesis. We emphasize that testing the Middle Eastern origin of European Jews can only be done with indigenous Middle Eastern groups. Overall, the similarity between European Jews and Caucasus populations underscores the genetic continuity that exists among Eurasian Jewish and non-Jewish Caucasus populations.

This genetic continuity is not surprising. The Caucasus gene pool proliferated from the Near Eastern pool due to an Upper Paleolithic (or Neolithic) migration and was shaped by significant genetic drift, due to relative isolation in the extremely mountainous landscape (Balanovsky et al. 2011; Pagani et al. 2011). Caucasus populations are therefore expected to be genetically distinct from Southern European and Middle Eastern populations (Figure 5), but share certain genetic similarity with Near Eastern populations such as Turks, Iranians, and Druze. This similarity, particularly with the Druze, should not be confused with a Semitic origin, which can be easily distinguished from the non-Semitic origin (Figure 5). In all our analyses, Middle Eastern samples clustered together or exhibited high similarity along a geographical gradient (Figure 3) and were distinguished from Arabian Peninsula Arab samples on one hand and from Near Eastern - Caucasus samples on the other hand.

Our study attempts to shed light on the Khazars and elucidate some of the most fascinating questions of their history. Although the Khazars' conversion to Judaism is not in dispute, there are questions as to how widespread and established the new religion became. Despite the limited sample size of European Jews, they represent members from the major residential Jewish countries (i.e., Poland and Germany) and exhibit very similar trends. Our findings support a large-scale conversion scenario that influenced the majority of the population. Another intriguing question touches upon the origins of the Khazars, speculated to be Turk, Tartar, or Mongol (Brook 2006). As expected from their common origin, Caucasus populations exhibit high genetic similarity to Iranian and Turks with mild Asian ancestry (Figure 5, Figure S7). However, we found a weak patrilineal Turkic contribution compared to Caucasus and Eastern European contributions (Figure 7). Our findings thus support the identification of Turks as the Caucasus ancestors, but not necessarily the predominant ancestors. Given their geographical position, it is likely the Khazarian gene pool was also influenced by Eastern European populations that are not represented in our dataset.

Our results fit with evidence from a wide range of fields. Linguistic findings depict Eastern European Jews as descended from a minority of Israelite-Palestinian Jewish emigrates who intermarried with a larger heterogeneous population of converts to Judaism from the Caucasus, the Balkans, and the Germano-Sorb lands (Wexler 1993). Yiddish, the language of Central and Eastern European Jews, began as a Slavic language that was re-lexified to High German at an early date (Wexler 1993). Our findings are also in agreement with genetic, archeological, historical, linguistic, and anthropological studies and reconcile contradicting genetic findings regarding European Jewish ancestry (Polak 1951; Patai and Patai 1975; Wexler 1993; Brook 2006; Kopelman et al. 2009; Sand 2009). Finally, our findings confirm both oral narratives and the canonical Jewish literature describing the Khazar's conversion to Judaism and the Judeo-Khazarian ancestry of European Jews (e.g., "Sefer ha-Ittim" by Rabbi Jehudah ben Barzillai [1100] , "Sefer ha-Kabbalah" by Abraham ben Daud [1161 CE], and "The Khazars" by Rabbi Jehudah Halevi [1140 CE]) (Polak 1951; Koestler 1976). We emphasize that we do not intend to cast doubt on Behar's et al. (2010). and Atzmon et al.'s (2010) remarkable findings, but rather propose a comprehensive interpretation that explains the patterns they observed in whole genome data, those reported in the literature for uniparental data, and those observed here using both types of data. The point in these studies is that European Jews had a single Middle Eastern origin is incomplete as neither study tested the Khazarian Hypothesis, to the extent done here. Last, although disease studies were not conducted using Caucasus and Near Eastern populations to the same extent as with European Jews (Chakravarti 2011), many diseases found in European Jews are also found in their ancestral groups in Southern Europe, the Caucasus, and the Near East, attesting to their complex origins (Ostrer 2001).

Because our study is the first to directly contrast the Rhineland and Khazarian Hypotheses, a caution is warranted in interpreting some of our results due to small sample sizes and availability of surrogate populations. To test the Khazarian Hypothesis, we used a crude model for the Khazar's population structure. Our admixture analysis suggests that certain ancestral elements in the Caucasus genetic pool were unique to the Khazars. Therefore, using few contemporary Caucasus populations as surrogates may capture only certain shades of the Khazarian genetic spectrum. Moreover, our conclusions regarding the fate of the Khazars are limited to European Jewish populations. Further studies may yield a more complex demographic model than the one tested here and illuminate the multi-ethnical population structure of the Khazars. Irrespective of these limitations, our results were robust across diverse types of analyses, and we hope that they will provide new perspectives for genetic, disease, medical, and anthropological studies.

**Conclusions**

We compared two genetic models for European Jewish ancestry depicting a mixed Khazarian-European-Middle Eastern and sole Middle Eastern origins. Contemporary populations were used as surrogate to the ancient Khazars and Judeans, and their relatedness to European Jews was compared over a comprehensive set of genetic analyses. Our findings support the Khazarian Hypothesis depicting a large Caucasus ancestry along with Southern European, Middle Eastern, and Eastern European ancestries, in agreement with recent studies and oral and written traditions. We conclude that the genome of European Jews is a tapestry of ancient populations including Judaized Khazars, Greco-Romans and Mesopotamian Jews, and Judeans and that their population structure was formed in the Caucasus and the banks of the Volga with roots stretching to Canaan and the banks of the Jordan.

**Supplemental Materials**

Supplementary Note

Supplementary figures

Table S1 – mT Haplogroups (table).

Table S2 – Y Haplogroups (table).

Table S3 – Studied Populations (table)

Table S4 – ASD analysis for Non-Jews (table)

Table S5 – ASD analysis for Jews (table)

Table S6 – Population Summary of ASD analysis (table)

Table S7 – ASD results (table)

All supplementary materials are available from the author's website at:

http://eelhaik.aravindachakravartilab.org/publications.html

**Acknowledgment**

I am grateful to Brian and Sharon Browning for their help with the IBD analysis and to my colleagues for their comments.

**Figures**

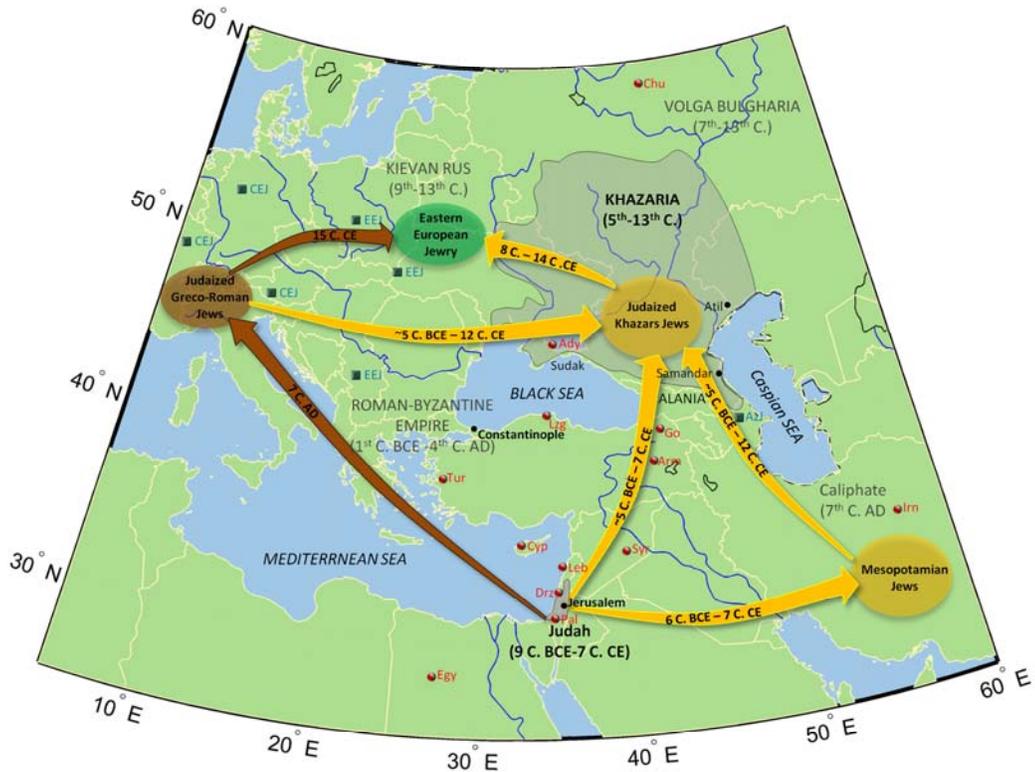

Figure 1. **Map of Eurasia.** A map of Khazaria and Judah is shown with the state of origin of the studied groups. Eurasian Jewish and non-Jewish populations used in all analyses are shown in square and round bullets, respectively (see Table S3). The major migrations that formed Eastern European Jewry according to the Khazarian and Rhineland Hypotheses are shown in yellow and brows, respectively.

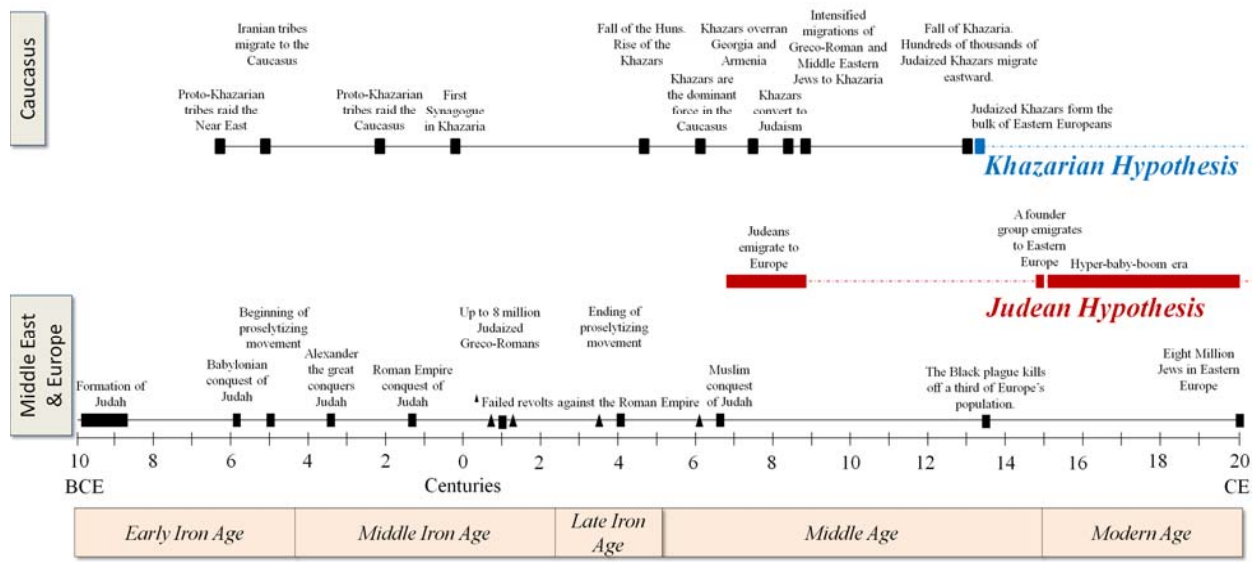

Figure 2. **An illustrated timeline for the relevant historical events**. The horizontal dashed lines represent controversial historical events explained by the different hypotheses, whereas solid black lines represent undisputed historical events.

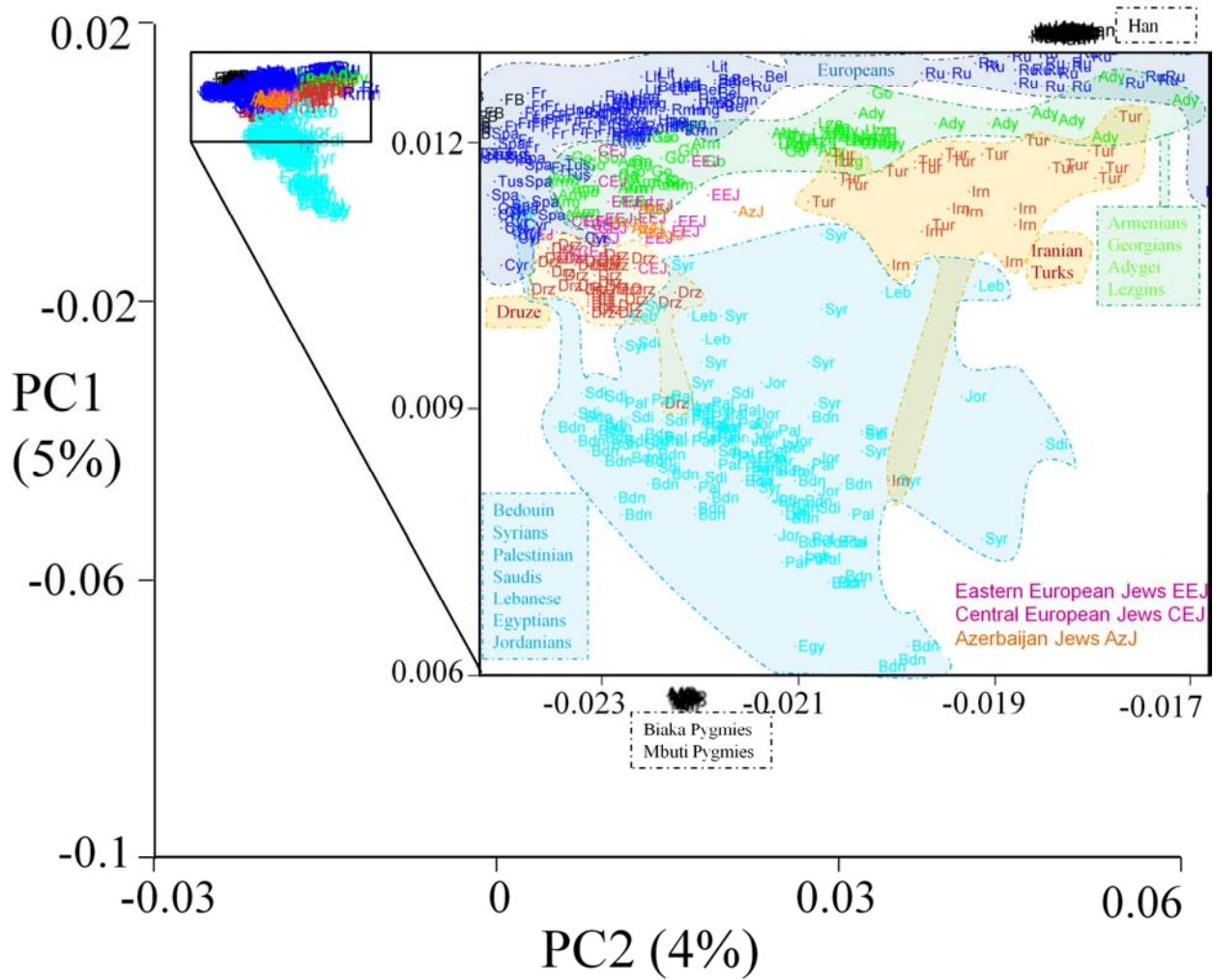

Figure 3. **Scatter plot of all populations along the first two principal components (Table S3).** For brevity, we show only the populations relevant to this study. The inset magnifies Eurasian and Middle Eastern individuals. Each letter code corresponds to one individual. A polygon surrounding all of the individual samples belonging to a group designation highlights several population groups.

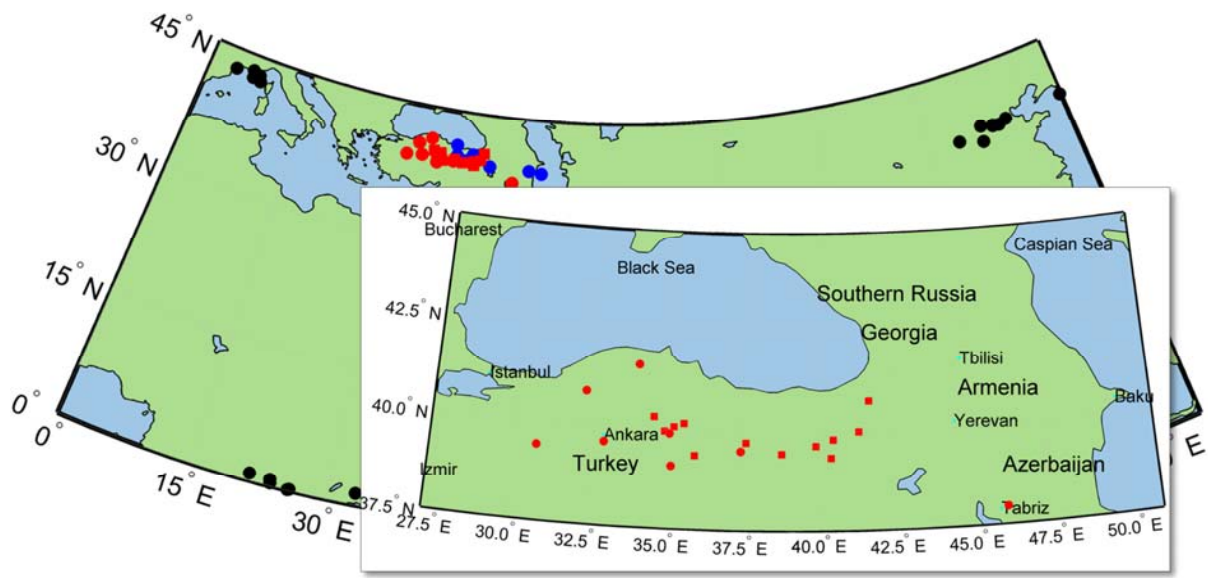

Figure 4. **Biogeographical origin of European Jews.** First two principal components were calculated for Pygmies, French Basque, Han Chinese (black), Armenians (blue), and Eastern or Central European Jews (red) - all of equal size. PCA was calculated separately for Eastern and Central European Jews and the results were merged. Using the first four populations as training set, Eastern (squares) and Central (circles) European Jews were assigned to geographical locations by fitting independent linear models for latitude and longitude as predicted by PC1 and PC2. Each shape represents an individual. Major cities are marked in cyan.

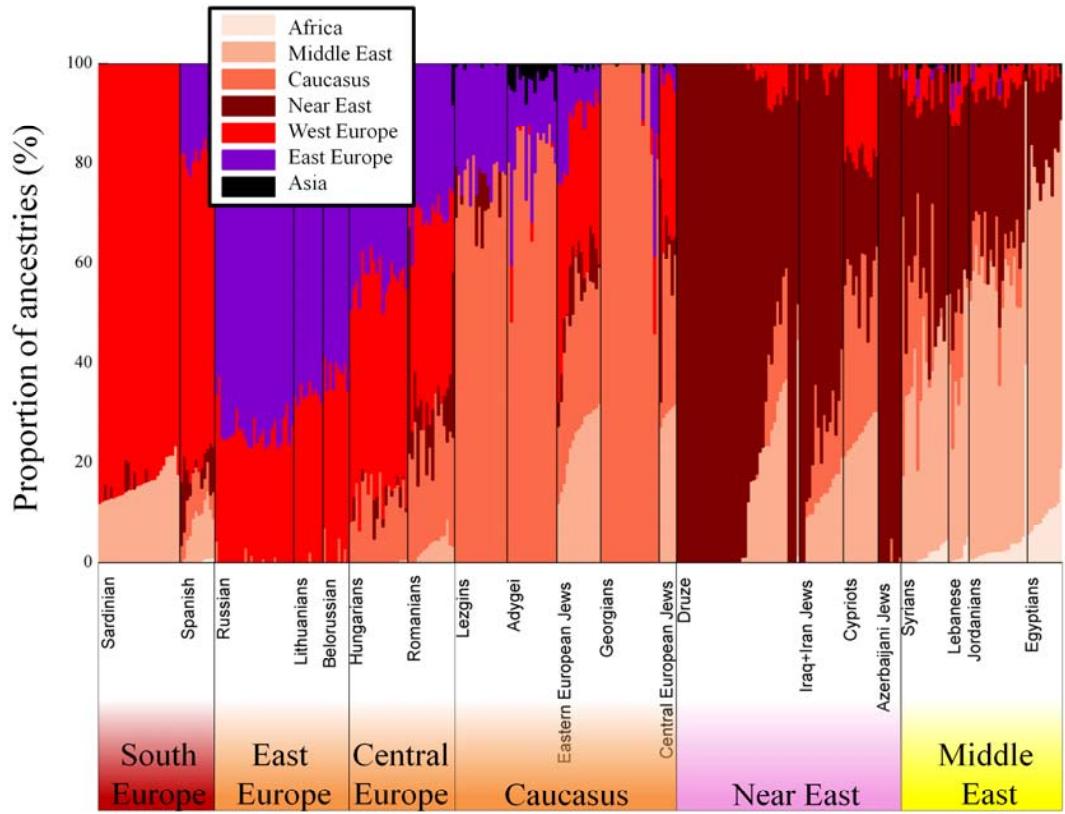

Figure 5. **Admixture analysis of Caucasus, Near Eastern, and Middle Eastern populations**. The *x*-axis represents individuals from populations sorted according to their ancestries and arrayed geographically roughly from North to South. Each individual is represented by a vertical stacked column (100%) of color-coded admixture proportions of the five ancestral populations. Here, the three Netherland Jews were grouped with Eastern European Jews due to their shared similarity.

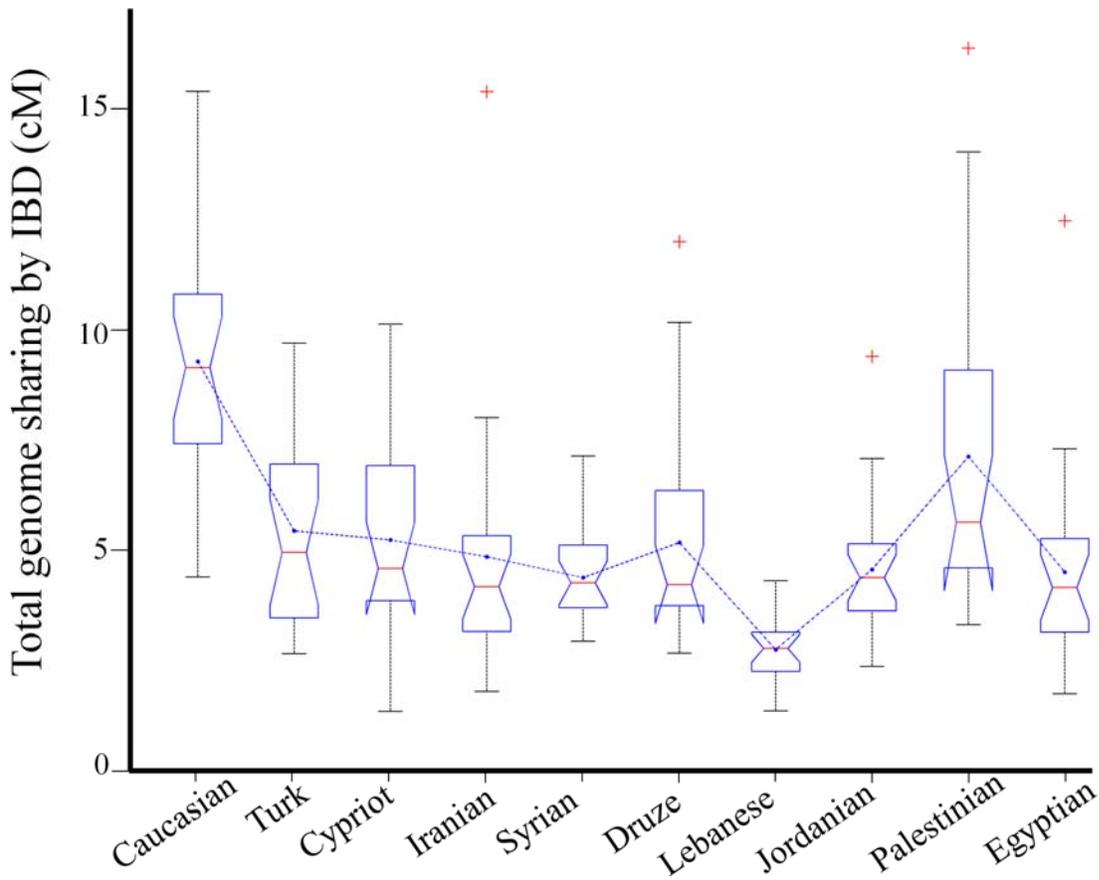

Figure 6. **Proportion of total IBD sharing between European Jews and different populations**. Populations are sorted by decreasing distance from the Caucasus. The maximal IBD between each European Jew and an individual from each population are summarized in box plots. Lines pass through the mean values.

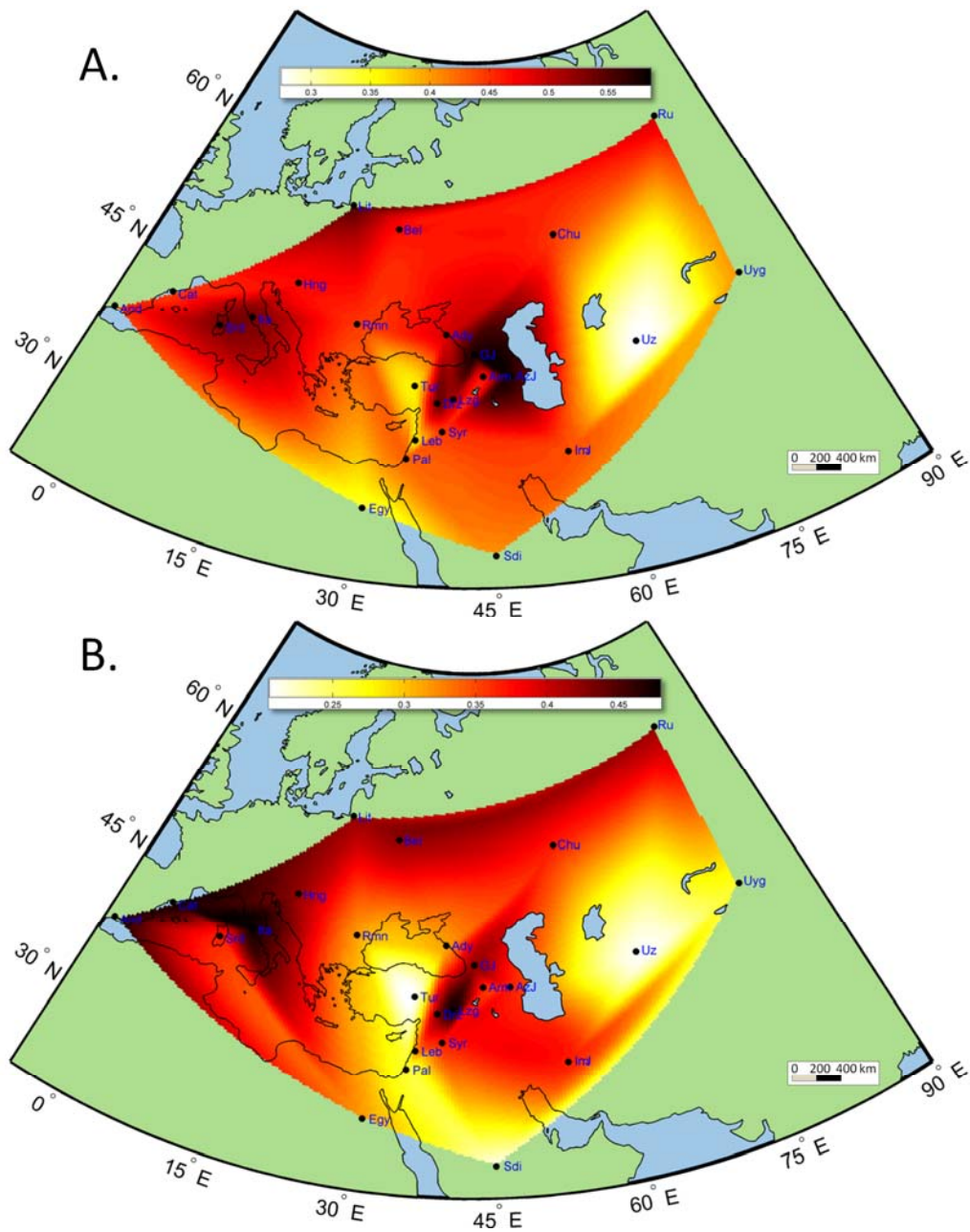

Figure 7. **Pairwise genetic distances between European Jews and other populations measured across a) mtDNA and b) Y-chromosomal haplogroup frequencies**. The values of $1-\delta_{xy}$ are color coded in a heat map with darker colors indicating higher haplogroup similarity with European Jews.

**Tables**

Table 1. **Genetic distances (ASD) between regional and continental Jewish communities (left panel) and between regional Jewish communities and their neighboring populations, Caucasus, and Middle Eastern population (right panel).** Bold entries are significantly smaller throughout each panel. The geographically nearest non-Jewish population were considered neighboring populations. The distances in the last two columns are between a Jewish and one Caucasus population (Armenians of Georgians) or Middle Eastern population (Palestinians, Bedouins, or Jordanians) that exhibited the lowest mean ASD.

| Regional Jewish community | Jewish populations | | | | Non-Jewish populations | | | |
|---|---|---|---|---|---|---|---|---|
| | Self | European | Asian | African | Neighboring population$^\Delta$ | | Caucasus$^\&$ | Middle Eastern$^\S$ |
| Eastern European | **0.2318** | 0.2328 | 0.2381 | 0.2446 | Hungarian | 0.2346 | **0.2340** | 0.2387 |
| Central European | **0.2312** | 0.2326 | 0.2378 | 0.2445 | Italians | **0.2335** | 0.2338 | 0.2385 |
| Bulgarian | 0.2326 | 0.2331 | 0.2376 | 0.2439 | Romanian | 0.2347 | **0.2337** | 0.2380 |
| Turkish | 0.2336 | 0.2336 | 0.2376 | 0.2439 | Turkish | 0.2353 | **0.2337** | 0.2379 |
| Iraqi | **0.2303** | 0.2351 | 0.2375 | 0.2447 | Iranian | 0.2363 | **0.2338** | 0.2381 |
| Georgian | 0.2304 | 0.2345 | 0.2372 | 0.2442 | Georgian | 0.2332 | **0.2332** | 0.2378 |
| Azerbaijani | **0.2304** | 0.2365 | 0.2386 | 0.2465 | Lezgins | 0.2367 | **0.2352** | 0.2398 |
| Iranian | **0.2310** | 0.2364 | 0.2391 | 0.2434 | Iranian | 0.2414 | **0.2361** | 0.2383 |